\documentclass{eas}
\usepackage{graphicx}
\newcommand\langeditorchanges[1]{{\bf \color{myred} #1}}

\usepackage{color}  
\definecolor{myred}{rgb}{0.7,0.0,0.2} 
\newcommand\edchange[1]{{\color{newred} #1}}

\usepackage{color}  
\definecolor{newred}{rgb}{0.7,0.0,0.2} 
\newcounter{exno}
\newcommand\ednote[1]{\addtocounter{exno}{1}%
\ifodd\value{page}
  \normalmarginpar
  \marginpar[#1]{\raggedright {\sf \color{blue} {\bf (\Roman{exno}}). #1}}
\else
  \reversemarginpar
  \marginpar[#1]{\raggedright {\sf \color{blue} {\bf (\Roman{exno}}). #1}}
\fi}
\renewcommand\langeditorchanges[1]{#1}

\renewcommand\edchange[1]{#1}
\renewcommand\ednote[1]{}
\begin{document}

\title{On the Super-Earths locked in the 3:2 \langeditorchanges{mean-motion} resonance} 
\runningtitle{\edchange{A. {\L}acny \& E.~Szuszkiewicz: Super-Earths locked in the 3:2 MMR}} 
\author{A.~{\L}acny}\address{Faculty of Mathematics and Physics, University of 
Szczecin  Wielkopolska 15, 70-451 Szczecin}
\author{E.~Szuszkiewicz}\address{CASA$^*$ and Institute of Physics, 
University of Szczecin, Wielkopolska 15, 70-451 Szczecin}
\begin{abstract}
  The first  study \langeditorchanges{of} migration-induced resonances in a pair of
  Earth-like planets has been performed by Papaloizou and Szuszkiewicz
  (\cite{ps}). \langeditorchanges{They} concluded that in \langeditorchanges{the} case of disparate masses embedded
  in a disc with the surface density expected for a  minimum mass solar
  nebula at 5.2 au, the most likely resonances are 
  \langeditorchanges{ratios of large integers, such as} 8:7. For equal masses, planets tend 
  to enter into the 2:1 or 3:2 resonance. In Papaloizou and Szuszkiewicz
  (\cite{ps}) the two low-mass planets have \langeditorchanges{masses} equal to 4 Earth 
  masses,
  chosen to mimic the very well known example of two  pulsar planets which
  are close to the 3:2 resonance. That study has stimulated quite a few
  interesting questions. One of them is considered here, namely
  how the behaviour of the planets close to the \langeditorchanges{mean-motion}
  resonance depends on the actual \langeditorchanges{values} of the \langeditorchanges{masses} of the planets. We have 
  chosen a 3:2 commensurability and investigated the outcome of an orbital
  migration in the vicinity of this resonance in the case of \langeditorchanges{a pair}
  of equal mass \langeditorchanges{super-Earths,} whose mass is \langeditorchanges{either 5 or 8}
  Earth \langeditorchanges{masses.} 

\end{abstract}
\maketitle
\section{Introduction and numerical set-up}
In the first study \langeditorchanges{of} migration-induced resonances in a pair of
low-mass planets Papaloizou and Szuszkiewicz (\cite{ps}) \langeditorchanges{performed}
simulations of two interacting
planets embedded in a disc with which they also interact (see also
Szuszkiewicz, this volume). The simulations
\langeditorchanges{were} carried out for a variety of planet masses, initial orbital
separations and initial surface densities in order to explore the possible
outcomes. In the case of planets with equal masses when the  relative
migration is slow, Papaloizou and Szuszkiewicz (\cite{ps}) have found that the 
planets become trapped
in the nearest available resonance.
In their \langeditorchanges{study,} the two low-mass planets \langeditorchanges{each have a mass of} 4 
Earth masses, chosen to mimic the very well known example of two  pulsar 
planets which are close to the 3:2 resonance. 
In this work we have modelled \langeditorchanges{the} evolution of \langeditorchanges{a pair} of planets
of equal mass in a gaseous protoplanetary \langeditorchanges{disc, again} in the vicinity of a 3:2
\langeditorchanges{mean-motion} resonance, but \langeditorchanges{for two different, larger, values of the mass, namely,} 5 and 8 Earth
\langeditorchanges{masses}.  

We have used the hydrodynamical code \langeditorchanges{NIRVANA,} originally written by
Ziegler (\cite{Ziegler}) and then adapted to disc-planet simulations. 
More details 
about the code can be found in the work by Nelson {\em et al.\/}
(\cite{nelson}).
The kinematic viscosity of the disk is zero and no magnetic field is present.  
The accretion of matter onto  planets is not included.
The units used in the code are the following. The mass of the central
object is taken to be the mass unit.
The initial distance $r_{p2}$ of the inner planet from the
central body  is \langeditorchanges{the} length unit. The time is
measured in 
      $\left(GM_{\odot}/{{r_{p2}}^3}\right)^{-1/2}$. 
The profile of the
initial surface density in the disc is essentially flat and it is described 
in \langeditorchanges{detail} in Papaloizou and Szuszkiewicz (\cite{ps}).  
The initial value of the surface density \langeditorchanges{$\Sigma_0$ at the locations of the planets} 
is $\Sigma_0=6.0\times 10^{-4}$ \langeditorchanges{in our units}.
The computational domain in polar coordinates \langeditorchanges{extends from 0.33 to 4}
length units in the radial direction and \langeditorchanges{covers} all azimuthal angles
from 0 to \langeditorchanges{2$\pi$, and} was divided into $384$
radial slices and $512$ angular sectors. The disc model is locally isothermal
with aspect ratio $H/r=0.05$

\section{Results}

In order to investigate how \langeditorchanges{orbital} evolution in the vicinity of a
3:2 resonance depends on \langeditorchanges{the} mass of the planets, 
we \langeditorchanges{first}  considered two planets \langeditorchanges{with} 5 Earth masses each and then
\langeditorchanges{we} repeated \langeditorchanges{the} calculation for two planets with 8 Earth masses each.
We assume that previous evolution brought the two planets to a configuration
close to the 3:2 commensurability. For the low-mass planets embedded
in a disc with the surface density distribution given in the form 
$\Sigma(r)\sim
r^{-\alpha}$ one can
estimate the characteristic time of migration $\tau$ on a circular orbit
from \langeditorchanges{the following} equation (see Tanaka, Takeuchi and Ward (\cite{tanaka}))
      \begin{equation} \label{eq:ttw}
      \tau=(2.7+1.1\alpha)^{-1}\frac{M_c}{M_p}\frac{M_c}{\Sigma_p r_p^2}
      \left(\frac{c}{r_p\Omega_p}\right)^2\Omega_p^{-1},  
      \end{equation}
\noindent where $M_c$ is the  mass of the central body, $M_p$ is the mass of 
the planet, $r_p$ is the semi-major axis of the orbit of the planet, 
$\Omega_p$ is the angular velocity of the planet
and $c$ is the sound speed in the disc at the distance $r_p$ from the
central body. 
We expect that each pair \langeditorchanges{of} planets will enter
the 3:2 mean motion resonance (MMR) because of their convergent  
migration, clearly following  from Equation (\ref{eq:ttw}). Among two 
planets with 
the same \langeditorchanges{mass,} the planet located further out from the central body will
migrate faster.

\begin{figure}
\centering
\vbox{
\hbox{ \hbox{\includegraphics[width=6cm,angle=0]{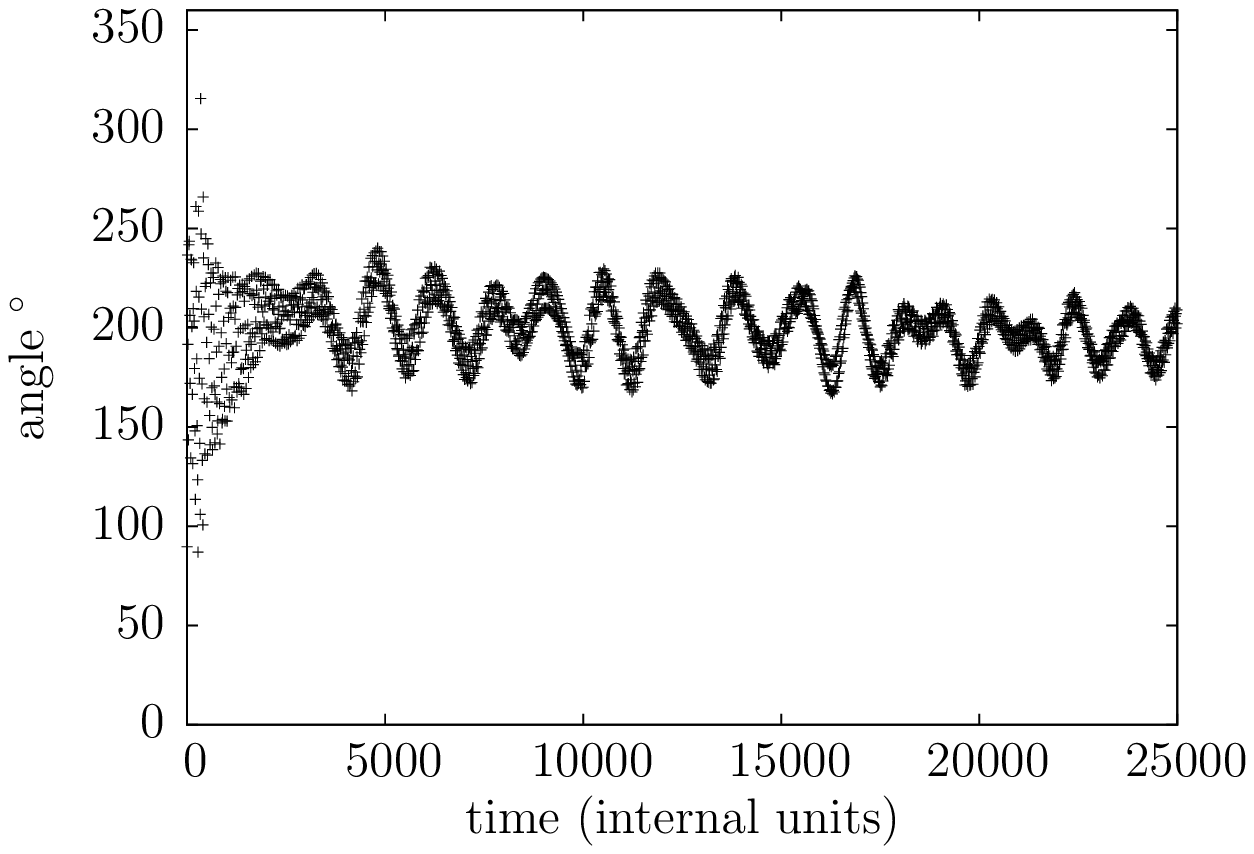}}
       \hbox{\includegraphics[width=6cm,angle=0]{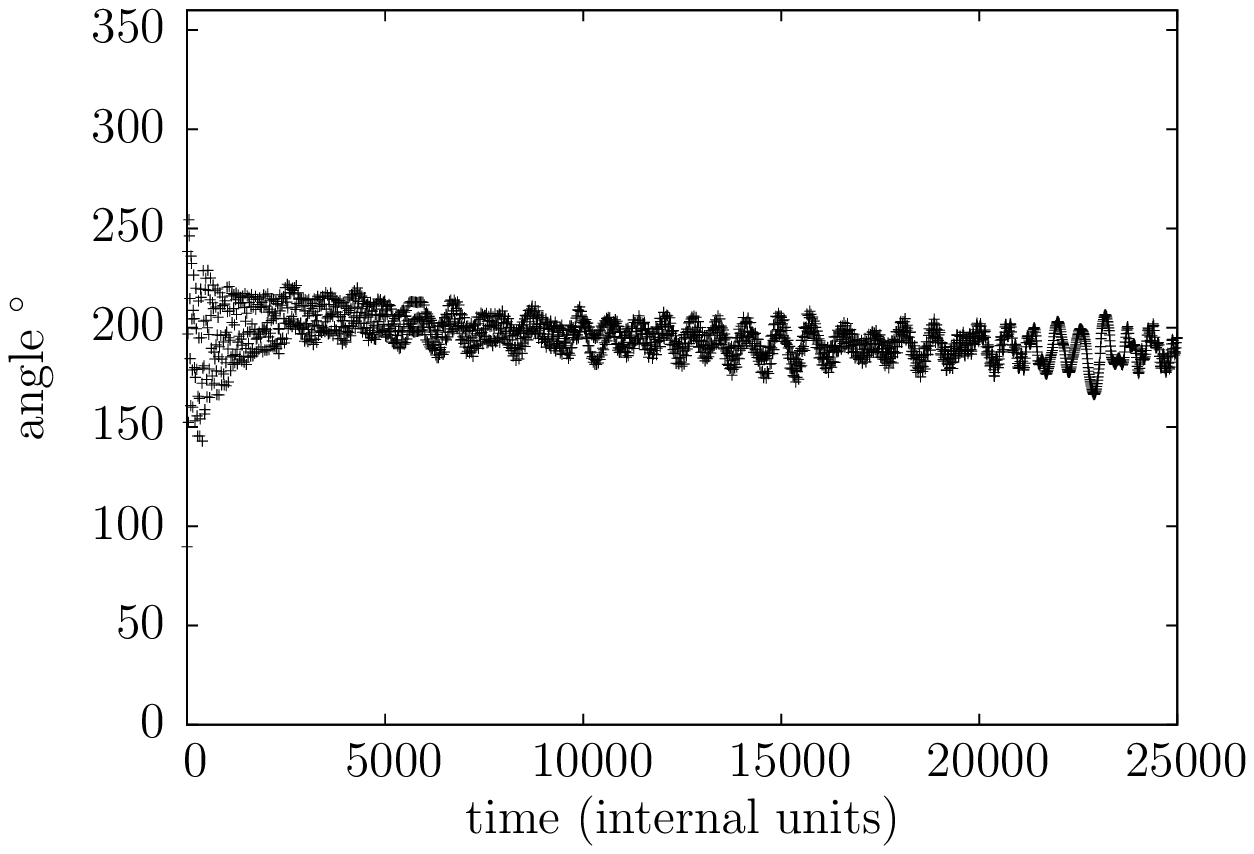}}
     }
\hbox{ \hbox{\includegraphics[width=6cm,angle=0]{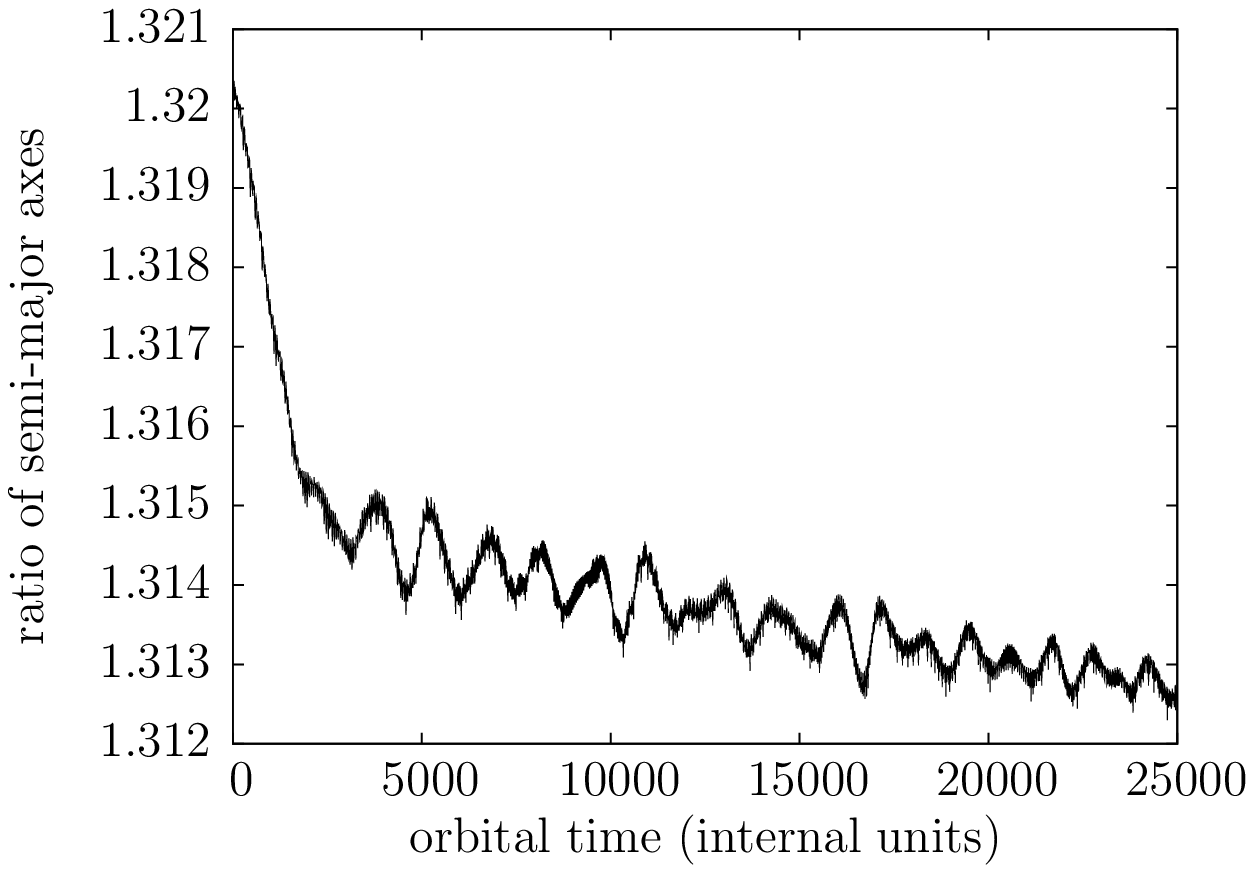}}
       \hbox{\includegraphics[width=6cm,angle=0]{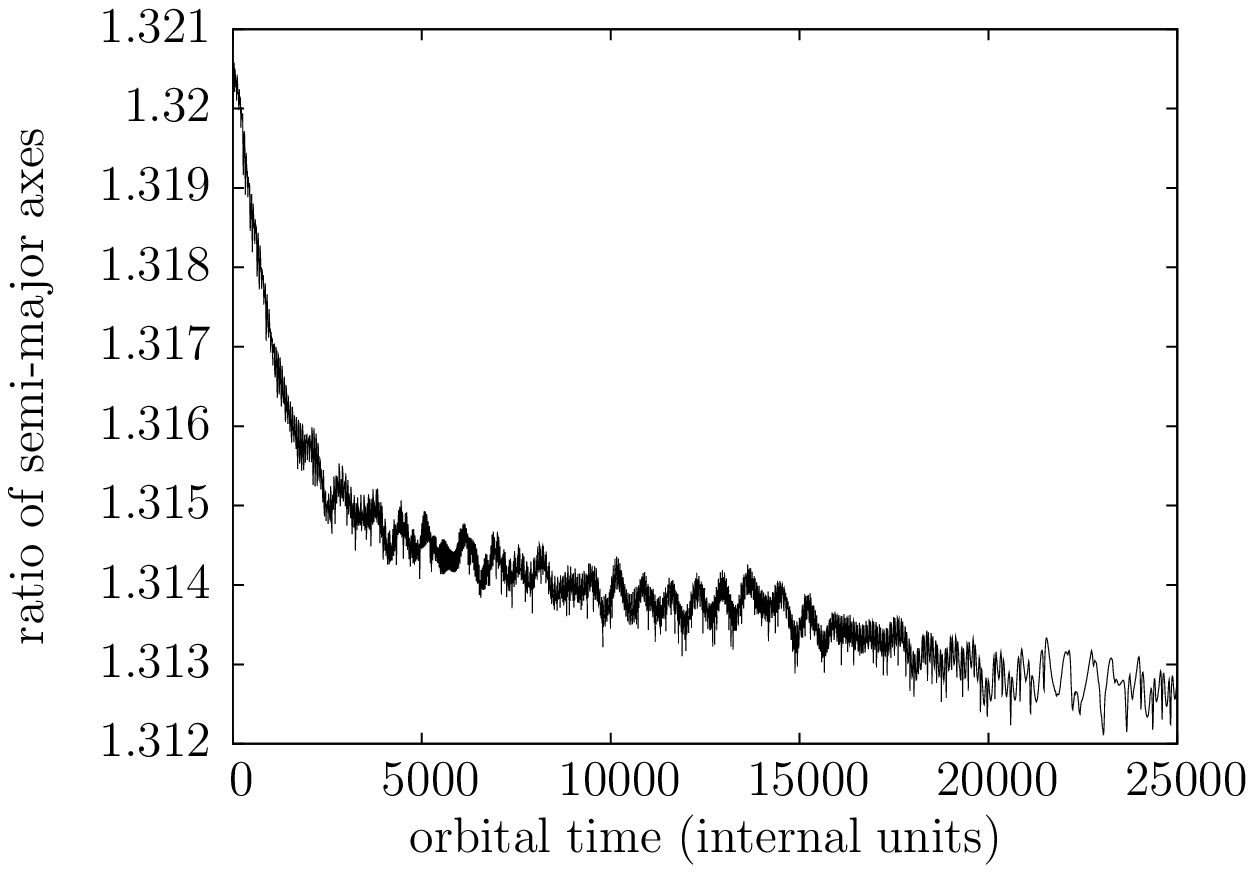}}
     }
\hbox{ \hbox{\includegraphics[width=6cm,angle=0]{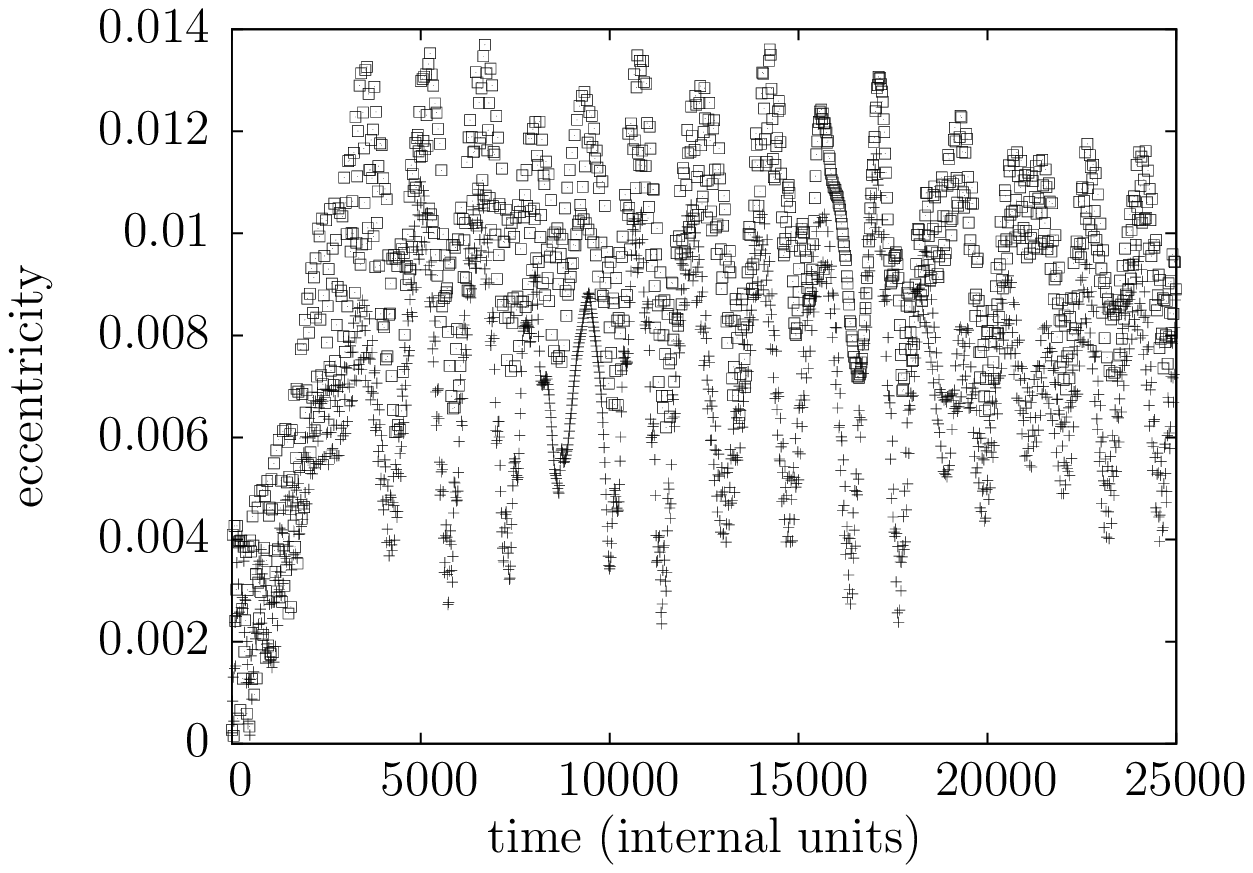}}
       \hbox{\includegraphics[width=6cm,angle=0]{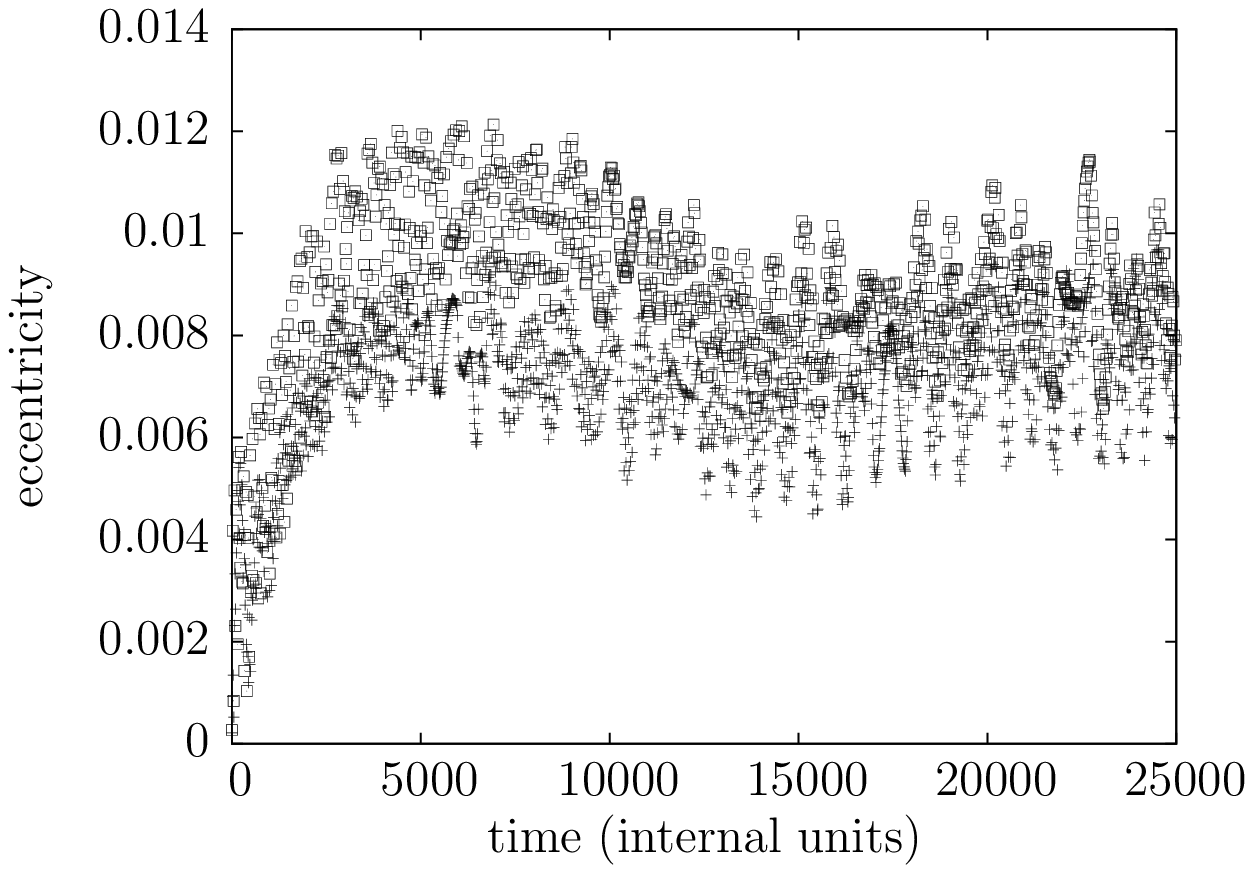}}
     }
}
	 \caption{\label{fig:m58}\langeditorchanges{Resonant} angles (upper panel), 
	 \langeditorchanges{ratio} of the semi-major axes
	 (middle panel) and \langeditorchanges{orbital} eccentricities (lower panel), 
	 crosses --
	 the inner planet, squares -- the outer planet, of the systems of 
	$5\mathrm{M}_\oplus$ planets -- left column, $
	8\mathrm{M}_\oplus$ planet \langeditorchanges{--} right column.
	 }
      \end{figure}

To analyse the system dynamics 
\langeditorchanges{close to the 3:2 
commensurability} 
in the presence of the disc 
we monitor the
evolution of the orbital elements, mainly the semi-major axes, eccentricities 
and
resonant angles defined as follows
      \begin{equation}\label{eq:phi1}
          \phi_i=3\lambda_1-2\lambda_2-\omega_i,
      \end{equation}
\noindent where $i=1,2$, $\lambda_i$ is 
      the mean longitude of the $i^{th}$ planet and $\omega_i$ is the
      longitude of the pericentre of the $i^{th}$ planet.
To identify the occurrence of the commensurability
we have used the fact that near the 3:2 resonance, the resonant angles
(\ref{eq:phi1}) and $\omega_1-\omega_2$ should librate about equilibrium 
values. 
In both cases the planets of $5\mathrm{M}_\oplus$ and $8\mathrm{M}_\oplus$
entered the 3:2 MMR and \langeditorchanges{stayed there throughout} the whole 
experiment from the beginning till the end, for 25000 dimensionless
time units \langeditorchanges{as} is seen in 
Figure~\ref{fig:m58}. 

\ednote{Please provide each panel of Fig.~1 in one .pdf or .eps file, 
to be included with includegraphics command, as shown below. The editorial office
requires that this command is used for the graphics inclusions.}

\section{Conclusions}
The planets with $5$ and $8\mathrm{M}_\oplus$
entered the resonance and
continued evolution in this state till the end of our experiment. 
The mean values of the orbital
eccentricities are excited to the value of $\sim 0.008$ in both cases and keep 
oscillating around this
value. 
\langeditorchanges{The} eccentricities of the internal and
external planets are roughly the same. The ratio of the orbital semi-major 
axes \langeditorchanges{slowly approaches} the
exact \langeditorchanges{resonance}. \langeditorchanges{A} full discussion of these results will appear
in \langeditorchanges{a} forthcoming paper.

\section{Acknowledgments}
This work has been partially supported by MNiSW grant N 203 026 32/3831
(2007-2010) and ASTROSIM-PL project. The simulations were done using Polish
National Cluster of Linux Systems (CLUSTERIX), the cluster HAL~9000 belonging
to The Faculty of Mathematics and Physics of The University of Szczecin and
computers provided by \edchange{Centrum Zarz\c{a}dzania Uczelnian\c{a}
Sieci\c{a} Komputerow\c{a} (CZUSK)}.  We want to thank the organizers of the
conference ``Extrasolar planets in multi-body systems: theory and
observations'' in Toru{\'n} (Poland), August 25-29, 2008, for very well
organized event. We are grateful to prof.  John Papaloizou for introducing us
into very interesting problem and for his advices, prof. Franco Ferrari for his
help in developing computational techniques necessary for success of our work,
and Edyta Podlewska for her valuable remarks and helpful discussion.


%
%
%
%
%
%
%

%
%
%


\ednote{Please check carefully the references part, it seems that is has been heavily
edited.}
\end{document}